\documentclass[12pt,letterpaper,aps,nofootinbib,floats,floatfix,amsmath,amssymb]{revtex4}
\usepackage{graphicx} \usepackage{hyperref} \usepackage{verbatim}
\begin{document}

\newcommand {\be} {\begin{equation}}
\newcommand {\ee} {\end{equation}}
\newcommand {\beae} {\begin{eqnarray}}
\newcommand {\eea} {\end{eqnarray}}
\newcommand {\eq} [1] {eq.\ (\ref{#1})}
\newcommand {\Eq} [1] {Eq.\ (\ref{#1})}
\newcommand {\eqs} [2] {eqs.\ (\ref{#1}) and (\ref{#2})}
\newcommand {\fig} [1] {fig.\ \ref{#1}}
\newcommand {\Fig} [1] {Fig.\ \ref{#1}}
\newcommand {\figs} [2] {figs.\ \ref{#1} and \ref{#2}}

\def\lsim{\mbox{\raisebox{-.6ex}{~$\stackrel{<}{\sim}$~}}}
\def\gsim{\mbox{\raisebox{-.6ex}{~$\stackrel{>}{\sim}$~}}}

\def \eps {\epsilon}
\def \veps {\varepsilon}
\def \pl {\partial}
\def \hf {{1 \over 2}}
\def \mf {\mathbf}
\def\figdir{}
\def\sss{\scriptscriptstyle}

\def\lsim{\mbox{\raisebox{-.6ex}{~$\stackrel{<}{\sim}$~}}}
\def\gsim{\mbox{\raisebox{-.6ex}{~$\stackrel{>}{\sim}$~}}}
\renewcommand{\t}{\tilde}

\noindent KCL-PH-TH/2011-{\bf 11}

\title{\bf Theoretical constraints on brane inflation and cosmic
  superstring radiation}

\author{Rhiannon Gwyn}
\email{rhiannon.gwyn@kcl.ac.uk}
\author{Mairi Sakellariadou}
\email{mairi.sakellariadou@kcl.ac.uk}

\author{Spyros Sypsas}
\email{spyridon.sypsas@kcl.ac.uk}
\affiliation{Department of Physics, King's College London,\\
Strand, London WC2R 2LS, U.K. }

\date{\today}

\begin{abstract}

We analyze theoretical constraints on the radiation modes of cosmic
superstrings. Given that cosmic superstrings are formed at the end of
brane inflation, we first investigate the implications of recently
elucidated supergravity constraints on brane inflation models.  We
show that both D3/D7 and D3/$\overline {\rm D3}$ brane inflation are
subject to non-trivial constraints.  Both inflationary models can be
shown to satisfy those constraints, but for the case of D3/D7 there
seem to be important consequences for the dynamics of the inflationary
mechanism.  Bearing this in mind, we analyze the theoretical
constraints on the nature of the allowed radiation by cosmic
superstrings in the context of a warped background where
brane-antibrane inflation takes place. Clearly such constraints do not
apply to field theoretic cosmic strings, or to cosmic strings that
arise in a background without warping.  We argue that in a warped
background where one might expect axionic radiation to be enhanced
relative to gravitational radiation, neither F-strings nor D-strings
can emit axionic radiation, and FD-strings cannot give rise to
Neveu-Schwarz--Neveu-Schwarz particle emission, while their
Ramond-Ramond particle emission is not well-defined.
\end{abstract}

\maketitle

\tableofcontents
\section{Introduction}
Cosmological inflation~\cite{infl} explains the large-scale
homogeneity, isotropy and flatness of the universe as observed
today. It also provides a model for structure formation (via
fluctuations of the inflaton field) whose predictions~\cite{infl-pert}
for the nature of the inhomogeneities in the Cosmic Microwave
Background (CMB) are in impressive agreement with
experiment~\cite{Komatsu:2010fb}. However, it is not known why
inflation occurred or which of the many models of inflation is
correct. Inflation often seems to require very special initial
conditions~\cite{generic}.  Theories of inflation, based on quantum
field theory combined with general relativity, can depend very
sensitively on Ultra-Violet (UV) physics in the sense that some
details of the inflationary dynamics are controlled by
Planck-suppressed contributions to the effective
action~\cite{str-th_infl}. One should therefore study inflation in a
UV-complete theory, such as string theory, in which one has the hope
of computing all Planck-suppressed contributions to the inflaton
action.

Although it arguably does not arise naturally, inflation can
be realized in various ways in string theory~\cite{str-th_infl,
  kallosh}, and can give rise to observational signatures, such as
deviations from scale invariance, gaussianity and adiabaticity of the
CMB. In particular, brane inflation models (in which the inflaton is
given by the separation between branes with an attractive potential
between them) generically results in the production of cosmic
superstrings. Observational consequences of cosmic superstrings can
serve as constraints on or signals of the underlying string theory
model~\cite{cs-review,Copeland:2009ga}.  Many studies have been done
on cosmic superstring networks, their evolution~\cite{cs-evol} and
radiation~\cite{cs-rad,Firouzjahi:2007dp}.  However, one must keep in
mind that in order for any predictions to be useful, they must be
consistent with the details of the string theory model under
consideration. In some cases, careful analysis reveals strong
constraints on the possible observational consequences of cosmic
superstrings from brane inflation. In what follows, we revisit these
models and their consequences, clarifying the possible signatures of
brane inflation via radiation from cosmic superstrings, within a
consistent theoretical framework.

This paper consists of two parts. In Section~\ref{inflation} we
examine the implications of recent supergravity constraints on brane
models of inflation, in which cosmic superstrings are
produced. Specifically we compare the well-known models of D3/D7
inflation and D3/$\overline {\rm D3}$ inflation. Since, as we will
show, D3/$\overline {\rm D3}$ inflation can be made compatible
with the supergravity constraints without spoiling the inflationary
dynamics, while for D3/D7 the D-term inflationary mechanism appears problematic,  in
Section~\ref{strings} we concentrate only on cosmic string radiation
in warped backgrounds in which brane-anti-brane inflation takes
place. Consistent compactification of such backgrounds leads to
further constraints on the allowed modes of radiation by these
strings. We end with our conclusions.

\section{Brane inflation within string theory}
\label{inflation}
In what follows we summarize the two classes of string theory
motivated brane inflationary models. We then examine the constraints
imposed by supergravity consistency conditions on these two classes
of models. The conclusions we will draw may lead to important
consequences for the feasibility of the models, and therefore the
phenomenology of cosmic superstrings formed at the end of brane inflation.

\subsection{Brane inflation models}
Brane inflation models in string theory fall into two classes:
\begin{itemize}
\item D3/D7 inflation~\cite{Dasgupta:2002ew, Dasgupta:2004dw} is a
  string realization of D-term inflation where the attraction between
  the branes is due to the breaking of supersymmetry by the presence
  of a non-self-dual flux on the D7-brane, which plays the r\^ole of a
  Fayet-Iliopoulos (FI) term. D3/D7 inflation takes place in an
  unwarped background: it is realized in type IIB string theory
  compactified on K3$\times {\rm T}^2/{\mathbb Z}_2$, where ${\mathbb
    Z}_2$ involves orbifold as well as orientifold operations. [We
    refer the reader to Refs.~\cite{Haack:2008yb, Burgess:2008ir}, for the details of
    the compactification including moduli stabilization.]
\item By contrast, brane-antibrane inflation~\cite{dvalitye}, of
  which D3/$\overline {\rm D3}$~\cite{Kachru:2003sx} is the best
  studied example, is a string realization of F-term inflation,
  where the attractive potential between the branes is due to
  warping. D3/$\overline{\rm D3}$ inflation takes place in a warped
  throat, such as the Klebanov-Strassler (KS)
  geometry~\cite{Klebanov:2000hb}, which is a deformed conifold
  geometry with fluxes. This brane inflation model necessitates a
  warped flux compactification~\cite{Dasgupta:1999ss, GKP}.
\end{itemize}
In both D3/D7 and D3/$\overline {\rm D3}$ models, flat directions for
the inflaton field can appear as a consequence of shift symmetry with
respect to the inflaton field.  

\subsection{Constraints from supergravity}
Several recent papers have addressed the consistency of having
constant (field-independent) FI terms in a supergravity
theory~\cite{FI-SUGRA,Komargodski:2010rb}.\footnote{These papers deal with the special case of supergravity theories obtained by coupling a rigid supersymmetric theory to supergravity, i.e. the case when the Bagger-Witten line bundle is trivial and the K\"ahler form exact. See e.g. \cite{Hellerman:2010fv} for further details.} Although there is no
problem with having such terms in a supersymmetric theory,
inconsistencies arise when such a theory is gauged. \footnote{ This holds in flat space. See Section \ref{braneantibrane} for a discussion of  the situation in AdS, studied in \cite{Adams:2011vw}.} These
inconsistencies can be understood in different
ways~\cite{FI-SUGRA,Komargodski:2010rb}, but in each case the
conclusion is the same: field-independent FI terms are inconsistent in
supergravity.

The same arguments rule out non-exact K\"ahler forms. In an equivalent
way, theories with a K\"ahler form that is not exact, or which have a
field-independent FI term, do not have a globally well-defined
Ferrara-Zumino (FZ) multiplet~\cite{FI-SUGRA,
  Komargodski:2010rb}.~\footnote{If the theory has a continuous
  R-symmetry, one can couple to supergravity using the so-called
  ${\cal R}$-multiplet.  However, in both cases --- gauging the theory
  by using either the FZ-multiplet or the ${\cal R}$-multiplet --- the
  resulting on-shell theory has a continuous global
  symmetry~\cite{FI-SUGRA}, and gravity theories with continuous
  global symmetries are expected to be inconsistent ({\sl see e.g.},
  Ref.~\cite{Banks:2010zn}, where it is argued that in quantum gravity
  models there are no global symmetries and all continuous gauge
  groups are compact).  When one has an FI term but no R-symmetry, one
  can couple to the S-multiplet, which interpolates between the FZ-
  and R-multiplets. The resulting supergravity theory contains an
  additional massless chiral superfield $\Phi$. It can equivalently be
  obtained by first adding such a field to the rigid theory so that
  the system including $\Phi$ has an FZ multiplet and can be coupled
  to supergravity as usual. This renders the FI term field-dependent
  and the moduli space non-compact~\cite{FI-SUGRA,
    Komargodski:2010rb}.\medskip}

To ensure that the FZ-multiplet exists and it remains well-defined,
the quantum moduli space must be such that the K\"ahler two-form $J$
is given by $d{\cal A}$, for some one-form ${\cal A}$ which is
globally well-defined. If this is the case, $\int J \wedge J \wedge J
\cdots$ vanishes for any compact cycle, giving a vanishing volume for
the moduli space of the internal manifold if it is compact. Thus, the
statement that the K\"ahler form must be exact is equivalent to the
statement that the moduli space cannot be
compact~\cite{Komargodski:2010rb}.~\footnote{This restriction on the
  moduli space is also indicated in earlier
  work~\cite{Witten:1982hu}.\medskip} In conclusion, the constraints
from supergravity are two-fold: 
\begin{itemize}
\item field-independent FI terms are not allowed
\item the moduli space cannot be compact.
\end{itemize}
In the following we investigate the implications of these (not
unrelated) conditions for the models of brane inflation discussed
above.

\subsubsection{D3/D7 inflation}
\label{D3D7}
D3/D7 inflation~\cite{Dasgupta:2002ew, Dasgupta:2004dw} is a stringy
realization of D-term inflation, a type of hybrid inflation.~\footnote
{The recently proposed model of flux-brane or D7/D7 brane inflation
  \cite{Hebecker:2011hk} is also a string embedding of D-term
  inflation and should be subject to the same general constraints, but
  we do not consider it here. An attractive property of this model is
  that cosmic strings produced at the end of flux-brane inflation are
  consistent with the observational bound on their tension, in
  contrast to D3/D7.}  In this model, a D3-brane is parallel to a
D7-brane in the four non-compact directions, with the other legs of
the D7 wrapping K3. The full compact manifold is K3$\times {\rm
  T}^2/{\mathbb Z}_2$. If there is a non-self-dual flux on the
D7-brane, the action of the system can be mapped to the action of a
D-term inflationary model in ${\cal N} = 1$ supergravity, with the
world-volume gauge field ${\cal F}$ playing the r\^ole of the FI term
(from the point of view of the field theory living on the brane).  An
attractive potential is then produced between the branes.  The
inflaton field is given by $\phi = x^4 + \imath x^5$ where $x^4$ and
$x^5$ are the directions perpendicular to both branes, along which
they feel the attractive potential. Inflation ends in the waterfall
stage, when strings stretching between the branes become
tachyonic. Supersymmetry is restored and $U(1)_{\rm FI}$ is
spontaneously broken in the final state, in which the D3-brane is
dissolved as an instanton into the
D7-brane~\cite{Dasgupta:2004dw}. Spontaneous symmetry breaking occurs
when the waterfall fields roll to a minimum at the end of this
process, which results in the production of cosmic
superstrings~\cite{Dasgupta:2004dw}. For D3/D7 inflation to take
place, it is necessary that the volume modulus be stabilized at some
large value for the K3 manifold. Failing to stabilize the
compactification modulus may result in rapid decompactification
instead of inflation.  Further, the effective gauge coupling on the D7
world-volume is given by $\tilde g_3$, where
\begin{eqnarray}
\frac{1}{\tilde g_3^2}  & = & \frac{{\rm vol}({\rm K3}) R^4 }{g_7^2},
\end{eqnarray}
and $R$ is an overall length scale of the K3. Thus it is inversely
proportional to the volume of the K3. This means that we can neglect
the interactions due to the D7 gauge fields as long as the K3
manifold is taken to be fixed at a large volume.  In addition, the
warp factor is of order 1 in this limit, allowing one to safely
ignore warping effects~\cite{Dasgupta:2004dw}.  
In conclusion, volume modulus stabilization allows one to ignore
warping effects and neglect interactions due to gauge fields from
any additional D7-branes, and ensures that the FI-term is non-zero,
which is necessary for D-term inflation to take place. The volume
modulus was left unfixed in Ref.~\cite{Dasgupta:2002ew,
  Dasgupta:2004dw} but was assumed to be large. In
Ref.~\cite{Haack:2008yb} the volume modulus was stabilized with a
non-perturbative superpotential due to gaugino condensation on a
stack of D7-branes wrapping the K3.

Let us investigate the impact of the supergravity constraints discussed above on D3/D7 inflation. In Ref.~\cite{Gwyn:2010rj} we
argued that the FI term in D3/D7, which arises because of a
non-self-dual flux on the D7 brane, is field-dependent and given by
\begin{eqnarray}
\label{FITERM}
\xi &= &\frac{\delta_{\rm GS}}{{\rm vol({\rm K3})}}~,
\end{eqnarray}
 where $\delta_{\rm GS}$ is the Green-Schwarz (GS)
 parameter~\cite{Green:1984sg}. This is fairly generic in string
 theory, where field-dependent FI terms arise from GS anomaly
 cancellation~\cite{Dine:1987xk,Dine:1987gj, Atick:1987gy}, and the
 connection between non-self-dual flux and a field-dependent FI term
 was alluded to in Refs.~\cite{Binetruy:2004hh, Burgess:2003ic}, where
 it was pointed out that the r\^{o}le of the axion in the GS
 mechanism in Type IIB will be played by the field dual to the
 4-form $C_{(4)}$, in the same multiplet as the volume modulus. Thus
 the r\^ole usually played by the axion-dilaton is played by the
 K\"ahler modulus $s = {\rm vol}({\rm K3})+\imath C_{(4)}$, giving the
 dependence of $\xi$ on ${\rm vol({\rm K3})}$.   

From this dependence, the first supergravity constraint ({\sl i.e.},
that  field-independent FI terms are not allowed) seems to be
satisfied, but for this to be true the real part of the K\"ahler
modulus, ${\rm vol(K3)}$, must be left unfixed. Specifically, the
volume of the K3 cannot be stabilized above the SUSY-breaking scale
in a consistent way~\cite{Komargodski:2010rb}. However, the SUSY
breaking scale in D3/D7 is given by $\xi${, which is tied to ${\rm vol(K3))}$: 
\begin{eqnarray*}
V & \sim & g^2 \xi^2 \,= \, g^2 \frac{\delta_{\rm GS}^2}{[\rm
  vol(K3)]^2}\, = \, \frac{\delta_{\rm GS}^2}{[\rm vol(K3)]^3}~,
\end{eqnarray*}
It is thus not possible to fix the volume below the SUSY breaking scale.}

Phrasing this another way, stabilizing the K\"ahler modulus at a large
finite value (as required for a successful model of inflation) above
the supersymmetry breaking scale would make the FI term constant, thus
problematic. Moreover this would amount to making the moduli space
compact, which is inconsistent with the second supergravity
constraint: In D3/D7 inflation, as in all proposed brane inflation
models, the moduli space of the effective world-volume theory is the
moduli space of the compactification manifold itself, fibered by the
K\"ahler moduli.  The open string moduli fields are the positions of
the mobile brane on the internal manifold that encode the geometry of
the compact base space while the closed string moduli are the K\"ahler
moduli of the manifold which encode the geometry of the fibration ,e.g. 
breathing modes of the internal manifold. These K\"ahler modes play an important r\^ole in the
model. Although the K3 (four-dimensional K\"ahler manifold) surface is
compact, the universal K\"ahler modulus, or volume modulus, fibers
over it yielding a non-compact moduli space, given by ${\mathbb R}^{+}
\times {\cal M}_{19,3}$~\cite{Becker:2007zj}, where the ${\mathbb
  R}^+$ factor corresponds to the overall volume modulus and the
${\cal M}_{19,3}$ factor describes a space of dimension $19\times
3=57$. If the volume modulus is fixed, the moduli space is rendered
compact. As opposed to the brane-anti-brane inflationary model to be
discussed in the next paragraph, the unavoidable constraint here is
the constant FI-term.

Given this analysis, one might conclude that as it stands D3/D7 is not
consistent with the supergravity constraints presented in
Ref.~\cite{FI-SUGRA, Komargodski:2010rb}. If the volume of K3 is
allowed to vary, we no longer have perturbative control of the theory,
do not necessarily get D-term inflation, and cannot neglect warping
with impunity. However, the distance $r$ between the branes is an
unfixed modulus in the theory, which the FI term $\xi$ appears to
depend on. There are two ways to see this, described in the following.

Firstly, the real part of the K\"ahler modulus, $s_R$, is not given
only by $\rm vol(K3)$ when quantum corrections are taken into
account. As explained in
Refs.~\cite{Haack:2008yb, Burgess:2008ir}, the physical warped
volume of K3 also depends on the D3 brane position, {\sl i.e.}, on the
brane separation $r$.~\footnote{Note that, as in
  Ref.~\cite{Gwyn:2010rj}, we use the notation of
  Ref.~\cite{Haack:2008yb} multiplied by a factor of $ \imath$ in
  order to be consistent with the notation of {\sl e.g.},
  Ref.~\cite{Davis:2005jf}.} This is in fact necessary in order to
avoid the so-called rho problem~\cite{Berg:2004ek, Giddings:2005ff,
  Baumann:2006th} and ensure that the definitions of the volume
modulus $\rho$ and of the D7 gauge coupling are consistent.  Using
Eq.~(\ref{FITERM}), this means that $\xi$ also depends on $r$, which
cannot be fixed as the branes move towards each other to give
inflation.

Secondly, one can find the $r$-dependence of $\xi$ directly from the
supergravity solution for the D3/D7 brane
system~\cite{Dasgupta:2002ew}. The FI term is given by the
non-self-dual part of the flux on the D7, namely
\cite{Dasgupta:2004dw}
\begin{eqnarray}
\frac{g^2 \xi^2}{2} & = & \frac{1}{8 R^{12} g_7^2} \int_{\rm K3} {\cal
  F}^- \wedge \star {\cal F}^-,
\end{eqnarray}
where the Hodge star is on the K3 wrapped by the D7 brane, $R$
represents the overall length scale of the compact K3, while $g$ and
$g_7$ are the $U(1)_{\rm FI}$ and D7 gauge coupling constants
respectively, and ${\cal F}^- = {\cal F} - \star {\cal F}$ where
${\cal F} = F - B$. One finds that
\begin{eqnarray}
\int_{\rm K3} {\cal F}^- \wedge \star {\cal F}^-& = & \int_{K3} d^4 x
\left [\frac{H_2}{H_1} (B_{67} - \sqrt{g} B_{89} )^2 + \frac{H_1}{H_2}
  (B_{89} - \sqrt{g} B_{67})^2 \right ]~,
\end{eqnarray}
using
\begin{eqnarray}
{\cal F}^- & = & - (B_{67} - \sqrt{g} B_{89}) dx^6 \wedge dx^7 -
(B_{89} - \sqrt{g} B_{67}) dx^8 \wedge dx^9~,\\ \sqrt{g} & = &
Z_7^{-1} H_1 H_2 ~,
\end{eqnarray}
with the B-field given by
\begin{eqnarray}
B_{67} & = & - \tan \theta_1 Z_7^{-1} H_1~,\\ B_{89} & = & - \tan
\theta _2 Z_7^{-1} H_2~,
\end{eqnarray}
where $H_i(r)$ are the harmonic functions corresponding to the branes,
and $Z_7$ is an r-dependent factor defined below. Neglecting terms of
order $(\sin \theta)^3$ and higher, as well as taking $r$ to be large,
we find
\begin{eqnarray}
\int_{\rm K3} {\cal F}^- \wedge \star {\cal F}^- &\approx &  \int d^4
x Z_7^{-3} (\sin \theta _1 - \sin \theta_2)^2~, 
\end{eqnarray}
where $Z_7 = 1 - 2 c_7 \ln \left(r/\Lambda \right ) $ gives the
$r$-dependence of this term. It is worth noting that if only the
leading order constant piece is identified with the FI term, namely
$\xi^2 \sim (\sin \theta_1 - \sin \theta_2)^2$, then the supergravity
constraint~\cite{FI-SUGRA, Komargodski:2010rb} applies, as the FI term
is now field-independent, or, alternatively, the moduli space is
compact. This implies theoretical inconsistency of the brane inflation
model. If $\xi$ depends on $r$, the FI term is no longer
field-independent and, furthermore, the SUSY breaking scale does not
depend only on ${\rm vol (K3)}$ so it may be possible to fix the
volume below the SUSY-breaking scale. However, the D-term inflationary
mechanism seems to be affected, in the sense that it is not
immediately clear how to interpret an $r$-dependent $\xi$ in D-term
inflation. More precisely, since the bifurcation point and the Hubble
constant during inflation both depend on $\xi$, an $r$-dependence of
the FI term would affect the inflationary mechanism.
\subsubsection{D3/$\overline {\rm D3}$ inflation}\label{D3D3bar}
\label{braneantibrane}
D3/$\overline {\rm D3}$ inflation is a stringy realization of F-
inflation, which is in general plagued by the $\eta$-problem. A
D3/$\overline {\rm D3}$ system lives at a specific point of a
Calabi-Yau (CY) manifold. Supersymmetry breaking results in a net
attractive force between the two branes, with the brane separation
playing the r\^ole of the inflaton. To accommodate sufficient
e-foldings, the first and second slow-roll parameters ($\epsilon$ and
$\eta$, respectively) must be small. However, since the separation
between the two branes cannot be greater than the size of the CY, one
cannot achieve $\eta\ll 1$ in flat space. To evade this problem, one
should consider the D3/$\overline {\rm D3}$ system in a warped
geometry~\cite{Kachru:2003sx}~\footnote{Alternative proposals that
  have been suggested are to consider branes at
  angles~\cite{Shandera:2003gx} or collisions of multiple branes. In
  the first case the number of e-foldings depends on the collision
  angle, while in the second one the slow-roll condition is
  abandoned.\medskip}.

In this model, usually referred to as $\mathbb{K}$L${\mathbb M}$T,
D3/$\overline{\rm D3}$ inflation takes place in a warped throat, in
contrast to the D3/D7 inflation case. The $\overline{\rm D3}$ brane is
fixed; it sits at the tip of the throat where its energy is minimized,
while the D3-brane feels a small attractive force towards the
$\overline{\rm D3}$ brane and moves towards it. In the
$\mathbb{K}$L${\mathbb M}$T model the inflaton field is given by 
separation between the D3 brane and the $\overline{\rm D3}$
brane. Inflation ends when the strings stretching between the branes
become tachyonic and the branes annihilate, with fundamental IIB
strings and D1-branes, localized in the throat, being naturally
produced~\cite{cs-prod}.  Due to the large gravitational red-shift at
the throat, the inflationary scale and the string tension measured by
a four-dimensional observer are suppressed by the warping factor, as
opposed to those measured by a ten-dimensional inertial observer which
are close to the four-dimensional Planck scale.~\footnote{Note that
  the original $\mathbb{K}$L${\mathbb M}$T model may be plagued by
  insufficient reheating. The U(1) gauge field on the stabilizing
  $\overline{\rm D3}$, is the only massless degree of freedom in the
  inflationary throat, and as such it is the only one that couples to
  the inflaton. As a result, at reheating almost all of the energy
  goes to the U(1) gauge bosons rather than into the Standard Model
  fields~\cite{Copeland:2009ga}.\medskip} 

Let us investigate whether D3/$\overline{\rm D3}$ is consistent with
the supergravity constraints put forward recently in
Refs.~\cite{FI-SUGRA, Komargodski:2010rb}. The theory has no FI term,
so it is only the second SUGRA constraint that we have to check,
namely that the moduli space of the theory is non-compact. 

 In the
compactifications of Refs.~\cite{Dasgupta:1999ss, GKP}, the volume
modulus is unfixed. However, in the D3/$\overline{\rm D3}$ inflation
model, the size of the internal manifold, which is a dynamical field,
can become large too fast and spoil the slow-roll
conditions~\cite{Kachru:2003sx}. Many efforts have been made to
stabilize the volume in order to solve this problem. More precisely,
one looks for a mechanism to stabilize the volume while leaving the
D3
 brane free to move in the CY.  Since the moduli space of the
theory is given by the manifold (in which the D3 brane is free to
move), fibered by the volume modulus
 which must be stabilized, one
might conclude that the moduli space is
 compact, making the model
inconsistent with the KS constraints. Specifically, it is not
consistent to stabilize the closed string modulus (the volume) at a
scale larger than the SUSY breaking scale ---  in other words, without SUSY breaking. It can be seen, however,
 that
$\mathbb{K}$L${\mathbb M}$T D3/$\overline{\rm D3}$ inflation
satisfies the supergravity consistency
constraints~\cite{Adams:2011vw}, as we  briefly
discuss below.

The supergravity constraints discussed in
Ref.~\cite{Komargodski:2010rb} hold for sigma
models in flat space, whereas D3/$\overline{\rm D3}$ inflation takes
place in a dS vacuum, as constructed in
KKLT~\cite{Kachru:2003aw}~\footnote{This was achieved by first
  constructing a SUSY-preserving AdS vacuum with all moduli fixed, and
  then adding a $\overline{\rm D3}$ brane which is fixed at the bottom
  of the throat, in order to lift the vacuum to dS (breaking
  SUSY).}. Recall that in flat space rigid ${\cal N} = 1$
supersymmetry does not impose any such conditions -- it is only when
SUSY is gauged that the constraints detailed by Komargodski and
Seiberg arise~\cite{Komargodski:2010rb}. However, as was recently
shown~\cite{Adams:2011vw}~\footnote{We thank McAllister for pointing
  us to Ref.~\cite{Adams:2011vw}.}, the KS conditions not
only extend to the case of $ {\cal N} = 1$ 
AdS$_4$ compactifications, but they in fact arise as
consistency conditions for unbroken ${\cal N} = 1$ SUSY on AdS$_4$
({\sl i.e.}, before gauging to supergravity). These conditions
follow from basic properties of SUSY in AdS$_4$.

The K\"ahler potential $K(\phi, \bar \phi)$ and the superpotential $W(\phi)$ are mixed by K\"ahler transformations 
\be
K(\phi, \bar \phi)\rightarrow K(\phi, \bar \phi)+f(\phi)+\bar f(\bar\phi)~~,~~
W(\phi)\rightarrow W(\phi)-\lambda f(\phi)~,
\nonumber\\
\ee 
in the AdS$_4$ case. Hence, requiring that the action be
invariant under K\"ahler transformations leads to a mixed potential of
the form~\cite{Adams:2011vw}:
\begin{eqnarray}
V(\phi, \bar \phi) & = & g^{i \bar j} (W_i + \lambda K_i)(\bar W_{\bar
  j} + \lambda K_{\bar j}) - 3 \lambda \bar W - 3 \lambda W - 3
\lambda^2 K~,
\end{eqnarray}
where $g^{i \bar j} W_i \bar W_{\bar j}$ is just the scalar
potential in flat space and $-3\lambda^2$ stands for the cosmological
constant. The mixing of the K\"ahler potential and superpotential in
rigid AdS$_4$ implies that, even when the superpotential is zero,
the SUSY vacua of this theory are generically a discrete set of
points, as opposed to the situation for flat-space sigma models where
the moduli space for vanishing superpotential is the full manifold.  For large volume flux compactification --- so that a
perturbation expansion in $\lambda/M_{\rm Pl}$ is valid --- to
AdS$_4$, with a small cosmological constant $\lambda$, the moduli
have masses proportional to $\lambda$, are therefore light. 
 However, if one wishes to consider this compactification in an inflationary
context one has to stabilize the K\"ahler moduli at a large value while keeping the open string moduli
(inflaton) nearly flat.

The authors of Ref.~\cite{Adams:2011vw} explicitly consider the
example of a theory with a mobile D3 brane and K\"ahler modulus, still
in the AdS$_4$ vacuum, {\sl i.e.}, before the addition of an anti-D3
brane as in KKLT~\cite{Kachru:2003aw}.  The K\"ahler potential is of
the form $K(\rho, \bar \rho, \phi, \bar \phi) = - 3 \log (\rho + \bar
\rho - k(\phi, \bar \phi))$.  Both $\rho$, the volume modulus, and $
\phi$, the D3-brane position, receive  Vacuum Expectation Values
(VEVs) with that of the volume still proportional to $\lambda$,
though parametrically larger than the AdS$_4$ scale. This still
corresponds to a moduli space composed of a set of discrete points, in
which inflation cannot happen. The moduli space spanned by $\rho$
and $\phi$, denoted $\hat X$, is consistent with the constraints
either when the volume is unfixed, giving a non-compact moduli space
$\hat X$, or when both the brane position and the volume take VEVs,
since a moduli space which is a discrete set of points has no compact
cycles. If $\rho$ becomes massive while the $\phi$ remain
massless, $\rho$ can be integrated out, leaving the compact moduli
space X, which is just the manifold. Since, as we have seen, this
is inconsistent, it is thus necessary to lift both the K\"ahler modulus
and the brane moduli, by breaking SUSY.

This is in fact the method proposed in $\mathbb{K}$L${\mathbb
  M}$T~\cite{Kachru:2003sx} for stabilizing the volume modulus while
leaving the brane position unfixed. We have seen that adding a
${\bar {\rm D}3}$ brane to break SUSY and give the D3 brane a
  nearly flat potential is insufficient if one wants to fix the
  volume. One can introduce a non-perturbative superpotential to
  stabilize the volume; although this generically gives rise to a
  large mass for the inflaton, spoiling inflation, it is possible to
  achieve inflation with a fine-tuned superpotential which also
  depends on the inflaton. \footnote{Many further attempts to stabilize the volume without ruining the flatness of the inflaton potential can be found in the literature, for instance in Refs.~\cite{Cline:2005ty, Baumann:2007ah}.} (This dependence is expected because the
  K\"ahler transformations away from flat space mix the K\"ahler 
  potential and the superpotential). Alternatively, the authors of
  Ref.~\cite{Kachru:2003sx} suggest stabilising $r \sim \rho + \bar
  \rho - k(\phi, \bar \phi)$ directly via corrections to the K\"ahler
  potential.

Both these methods involve SUSY breaking, so that beneath the SUSY
scale one can adjust the volume modulus as necessary, while contriving
to safeguard the inflationary behaviour of the system: this means that
in the absence of interactions the D3 brane should be free to move
around the compact CY. The crucial difference between this case and
that of the D3/D7 system in this regard is that the SUSY breaking
scale for D3/$\overline{\rm D3}$ is not tied to the scale of the
volume. This makes it possible to circumvent the consistency
constraints by fixing the volume {\em below} the SUSY breaking scale,
which is not possible for D3/D7 with FI term given by
Eq.~(\ref{FITERM}),  as we have seen. Thus, although it may require fine tuning to
pull it off, a D3/$\overline{\rm D3}$ inflationary scenario with
stabilized volume can be consistent with the constraints discussed
in Refs.~\cite{Komargodski:2010rb,Adams:2011vw}.
\subsection{Summary}
We have seen that the supergravity consistency conditions detailed in
Refs.~\cite{FI-SUGRA, Komargodski:2010rb} impose stringent constraints
on models of brane inflation in string theory. Assuming that these
constraints hold in the full string theory~\footnote{This assumption
  may turn out to be false. Supergravity is a low-energy description
  of string theory, which does not include $\alpha^\prime$ or higher
  derivative corrections. Thus, it is conceivable that the
  constraints from supergravity are weakened when these corrections
  are taken into account, and/or the brane inflation models need
  modifications from their current description.}, we are forced to do
a careful analysis of the details of compactification and moduli
stabilization in brane inflation models to ensure they are consistent
with these constraints.  

Most generally, as explained in Ref.~\cite{Komargodski:2010rb}, it is
difficult to allow open string degrees of freedom to remain massless
while stabilizing closed string degrees of freedom, such as the volume
of the compactification manifold. This can generally lead to a compact
moduli space for the theory, which is inconsistent~\cite{Komargodski:2010rb}.
However, as we argued above, this problem can be evaded in the case of
D3/$\overline{\rm D3}$.  

In the presence of an FI term, as in D-term models such as D3/D7, the
story is more complicated. Once again it is necessary and difficult
to stabilize the volume of the compact manifold, and furthermore
stabilizing the volume of K3 would make the FI term, at first glance
proportional to
 $1/{\rm vol (K3)}$, field-independent and therefore
the brane inflation model would be inconsistent. The alternative,
leaving the volume unfixed, means that inflation may be difficult to
achieve, and one can neither neglect warping nor gauge interactions
from additional D7 branes. These difficulties are avoided if, as
argued above, $\xi$ depends on the unfixed modulus $r$, the brane
separation. However, it is not immediately clear whether inflation
can proceed as usual if this dependence is taken into account.

\section{Cosmic superstring radiation in warped backgrounds}
\label{strings}
\subsection{Cosmic superstrings}
Cosmic strings are one-dimensional topological defects produced
generically~\cite{Jeannerot:2003qv} during phase transitions in the
early universe~\cite{cs-form}, whenever the resulting vacuum manifold
$\cal{M}$ has non-trivial first homotopy group $\pi_1({\cal
  M})$. Because they carry energy, cosmic strings can seed
gravitational instabilities and were therefore investigated throughout
the 80s and 90s as sources for structure formation. In particular,
GUT-scale cosmic strings lead to an adiabatic spectrum of primordial
density fluctuations, in agreement with COBE-DMR
measurements. However, data from BOOMERanG, DASI, MAXIMA and SASKATOON
in the late 90s, and more recently data from WMAP, showed that
inflation, which gives rise to an adiabatic scale-invariant power
spectrum with Gaussian statistics, must be the dominant source of
primordial density fluctuations, with cosmic strings giving a
sub-dominant (but not negligible) contribution~\cite{Bouchet:2000hd}.
Despite the fact that GUT-scale cosmic strings are not the dominant
source of primordial density perturbations, data is consistent with
their sourcing up to 10 \% of these perturbations and with a string
tension satisfying the bound~\cite{bound}
\begin{equation}
\label{tensionbound}
G \mu \leq 2 \times 10^{-7}~,
\end{equation}
where $G$ is Newton's constant.  These strings could be observed via
the gravitational radiation they emit or via gravitational lensing.

Interest in cosmic strings then waned until it was shown that cosmic
superstrings are generically produced at the end of brane
inflation~\cite{cs-prod} and could provide an observational window on
string theory. In contrast to their gauge theory analogues, cosmic
superstrings in general have Planck-scale energies, leading to $G \mu
\sim 1$, an inconsistently high tension. These objects are therefore
undesirable in string theories of the early universe, and models in
which they are produced and remain stable must be ruled out.  In a
warped geometry~\footnote{The tension can also be lowered in models
  with large extra dimensions, which we do not consider here.},
however, such as in the case of the throat in which brane-antibrane
inflation takes place, the tension of cosmic superstrings can easily
be lowered to within the bound given in Eq.~(\ref{tensionbound})
above~\cite{Firouzjahi:2005dh}.  In the case of D3/D7 inflation, the
problem of an unacceptably high string tension can be solved by making
the strings semi-local~\cite{Dasgupta:2004dw}, since the upper bound
on the tension of semi-local strings is higher than that for local
abelian strings, or by suppressing the string production by taking
higher order corrections to the K\"ahler potential into account, as in
Ref.~\cite{Haack:2008yb}.

Cosmic superstrings, produced at the end of brane
inflation~\cite{cs-prod}, can be F-strings (the fundamental strings of
string theory), D-strings (one-dimensional Dirichlet branes) or
$(p,q)$ strings, which are bound states of $p$ F-strings and $q$
D-strings.  Stable or metastable strings arising from wrapped
D-branes, Neveu-Schwarz (NS) branes and M-branes have also been
constructed~\cite{CMP,Leblond:2004uc}; these are higher-dimensional
branes that are wrapped on compact cycles resulting in only one
remaining non-compact dimension.  Thus, the cosmic superstrings of
string theory come in many more varieties than those of a GUT and with
different interactions and a richer
spectrum~\cite{CMP,cs-branes}. This realization opened the possibility
of a new observational window on string theory models and led to a
renewal of interest in the subject. It is our aim to explore this
window via radiation from cosmic superstrings. Taking a consistent
compactification into account imposes certain constraints on the
allowed radiation from these strings, as we will see below. Having
examined possible constraints on the brane inflation models in which
cosmic superstrings can be produced, we proceed with the allowed
channels of cosmic superstring radiation in warped backgrounds.

\subsection{Warped compactifications}
\label{warping}

In a warped geometry,  the metric takes the form
\begin{eqnarray}
\label{warped}
ds^2 & = & e^{2 A(y)} g _ {\mu \nu} dx^\mu dx^\nu + e^{- 2 A(y)} g_{mn}dy^m
dy^n~,
\end{eqnarray}
where $e^{2A(y)}$ is the warp factor. Here and in the rest of the
paper greek indices run over the $0,1,2,3$ directions; $y^m, y^n,...$
denote the internal directions. The internal metric includes a
throat-like region such as the warped deformed conifold of the
Klebanov-Strassler geometry~\cite{Klebanov:2000hb}, for which the
metric and fluxes are well known. In the warped deformed case the
conifold singularity is smoothed away by fluxes.

Such warping can be produced by a stack of D3-branes (as in the
AdS/CFT correspondence~\cite{ads}; AdS$_5$ can be represented as a
four-dimensional space plus radial direction with warp factor) and
provides a way to obtain hierarchies of scale in four
dimensions~\cite{Randall:1999ee}. However, the warped KS-type throat
in which brane-antibrane inflation takes place is non-compact. In
order to include it in a string theory setting, the throat must be
``glued" to a compact geometry which gives rise to reasonable
four-dimensional physics ({\sl e.g.}, a finite Planck scale in four
dimensions). This is achieved using negative tension objects such as
orientifold planes, which are needed to satisfy the integrated field
equations and result in stringent constraints on the fields and fluxes
in the theory~\cite{Dasgupta:1999ss, GKP}. Compactifying on an
orientifold results in a truncated spectrum: the orientifold action
projects out certain fields~\cite{Grana:2003ek, Grimm:2004uq}.

In addition, the dimensional reduction of different light fields is
non-trivial in flux compactifications~\cite{Frey:2008xw}. For
instance, consider the universal volume, or K\"ahler modulus, which in
the unwarped case corresponds to a rescaling
\begin{eqnarray}
g_{mn} \rightarrow e^{2 u} g_{mn}~,
\end{eqnarray}
and fluctuation
\begin{eqnarray}
ds^2 & = & e^{-6 u(x)} g_{\mu \nu} dx^\mu dx^\nu~,
\end{eqnarray}
and which pairs with the universal axion given by
\begin{eqnarray}
C_4 & = & \frac{1}{2} a(x) J \wedge J~,
\end{eqnarray}
where $J$ is the K\"ahler form of the CY, into the complex field $\rho
= a + \imath e^{4 u}$. In a warped background of the form
Eq.~(\ref{warped}), it is not immediately clear how to define the
fluctuations $u$ or $a$. Naive attempts for $u$, such as writing
\begin{eqnarray}
ds^2 & = & e^{2 A(y)} e^{-6 u(x)}\eta_{\mu \nu} dx^\mu dx^\nu + e^{- 2
  A(y) }e^{2u(x)} g_{mn} (y) dy^m dy^n~,
\end{eqnarray}
do not solve the ten-dimensional Einstein equations. It turns out that
additional components (called compensators) in the metric will be
required~\cite{Frey:2008xw}, complicating the dimensional
reduction. Similarly, (two) compensators are required for 
definition of the universal axion, and these enter in the equations of
motion while obeying defining constraints. Thus, the universal K\"ahler
modulus and universal axion wave-functions can be given, at least
formally. The same is not true in the case of the non-universal axion
which, as we shall see, is the only possible mode for axion emission
from $(p,q)$ strings in a throat.

\subsection{Allowed radiation from cosmic superstrings in a throat}
We are interested in the types and signatures of radiation from 
superstring networks. For cosmic string networks to reach a scaling
solution, in which the characteristic length scale of the network
scales with time, the networks must lose energy. For conventional
gauge strings, the dominant mode of this energy loss occurs when loops
radiate away into gravitational radiation. Gravitational radiation is
also possible (but sub-dominant) for other processes in the evolution
of these networks, such as reconnection.

For cosmic superstrings, which are charged under the two-forms $B_2$
and $C_2$, axionic radiation is also possible, because these two-forms
are Hodge dual to axionic scalars in four dimensions. In the case of
D-strings, it has been argued~\cite{Firouzjahi:2007dp} that axionic
radiation can be the dominant mode of energy loss in a warped
geometry, because the warp factor does not couple to the Chern-Simons
part of the action, in which $C_2$ appears.

It might seem that this result would translate to the case of $(p,q)$
strings in a warped background, which is where they are expected to be
produced at the end of brane-antibrane inflation. This would give a
clear observational difference between cosmic superstrings and gauge
theory cosmic strings.  However, taking the details of the string
compactification into account leads to difficulties with the argument
in Ref.~\cite{Firouzjahi:2007dp}. More precisely, $C_2$ and $B_2$ are
projected out of the spectrum, and the axion wave-function is
non-trivially modified by the warp factor. We find that there is no
well-defined axionic mode for the $(p,q)$ strings in a throat.

\subsection{Gravitational radiation}
A $(p,q)$ string in a KS throat can be constructed by wrapping a
D3-brane with suitable charges on a 2-cycle (which is stabilized by
fluxes)~\cite{Firouzjahi:2006vp}. The action is given by
\begin{eqnarray}
S_{\rm D3} & = & - \frac{\mu_3}{g_s} \int d^4 x \sqrt{- |g_{ab} + {\cal
    F}_{ab}}| + \mu_3 \int \left ( C_2 \wedge {\cal F} + \frac{1}{2}
C_0 {\cal F} \wedge {\cal F}\right )~,
\end{eqnarray}
where $\mu_3$ is the D3-brane tension, the integral and $a,b$ indices
run over the four-dimensional world-volume $0,1,2,3$, where $2$ and
$3$ are the coordinates on the 2-cycle the D3-brane wraps, and ${\cal
  F}_{ab} = B_{ab} + \lambda F_{ab}$, where $\lambda = 2\pi\alpha'$
and $\mu_3= 1/(2\pi)^3 \alpha'^2$.

Here the metric reads
\begin{eqnarray}
ds^2 & = & h^2 \eta_{\mu \nu} dx^\mu dx^\nu + ds_6^2~,
\end{eqnarray}
where $h$ is the warp factor (which is a function of the internal
coordinates and has thus been absorbed into $ds_6^2$), and the
necessary fluxes are
\begin{eqnarray}
F_{23} & = & \frac{q}{2} \sin \theta d \theta d \phi~,\nonumber\\
\tilde F^{01} & = & - \frac{p}{4 \pi}~,\nonumber\\
B_{23} &\neq & 0,
\end{eqnarray}
where $\tilde F^{\mu\nu}$ denotes the conjugate of the electric
field. Hence, we can find the stress-energy tensor $T^{\mu \nu}_{\rm
  DBI}$ derived from the Dirac-Born-Infeld (DBI) action:
  \begin{eqnarray}
       T^{\mu \nu}_{\rm DBI} & = & - \int d^2 \zeta \frac{\mu_3}{g_s}
       \frac{(g_{22} g_{33} + {\cal F}_{23}^2)^{1/2}}{(h^4 - \lambda^2
         F_{10}^2)^{1/2}} h^4 [\dot X^\mu \dot X^\nu - X^{\mu '}
         X^{\nu '}]~. 
  \end{eqnarray}
Upon minimization, the (00)-component
  \begin{eqnarray}
  T^{00} & = & \int d^2 \zeta \frac{\mu_3}{g_s} \frac{(g_{22} g_{33} +
    {\cal F}_{23}^2)^{1/2}}{(h^4 - \lambda^2 F_{10}^2)^{1/2}} h^4~,
    \end{eqnarray}
leads to:
\begin{eqnarray}
T_{(p,q)} & = & \frac{h_I^2}{2 \pi \alpha'} \sqrt{\frac{q^2}{g_s^2} +
  \left (\frac{bM}{\pi} \right )^2 \sin^2 \left (\frac{\pi (p - q
    C_0)}{M} \right ) }\nonumber\\ & = & \sqrt{T_{\rm D}^2 + T_{\rm F}^2}~,
\label{pq-gr-rad}
\end{eqnarray}
with $T_{\rm D}, T_{\rm F}$ denoting the tensions of the D-string and
F-string, respectively. This reduces to the flat space expression
$T_{(p,q)} = T_{F1} \sqrt{(q^2/g_s^2) +
  p^2}$~\cite{Schwarz:1995dk} when $b = h_I = 1$ (and $C_0 =
0$)~\cite{Firouzjahi:2006vp}. Thus, the DBI part of the action has the
same effect as the usual Nambu-Goto action, except that the tension
of the string is modified. Similarly, the string equation of motion
arises from varying the action with respect to $\delta X^\mu$, so
neither the Chern-Simons terms nor the ${\cal F}_{ab}$ factor will
contribute to these equations.  Given $T^{\mu \nu}$, the
gravitational radiation from $(p,q)$ strings is obtained from
\begin{eqnarray}
\frac{d P_n}{d \Omega} & = & \frac{G \omega_n^2}{\Pi} [T_{\mu \nu}^
  \star (\omega_n, \vec{k}) T^{\mu \nu} (\omega_n, \vec{k}) -
  \frac{1}{2} |T^\nu _\nu ( \omega_n, \vec{k})|^2]~,\\ 
\mbox{where}\ \ \ \ \ \  T_{\mu \nu}
(\omega_n, \vec{k}) & = & \frac{2}{L} \int^{L/2}_0 dt e^{\imath
  \omega_n t} \int d^3 x e^{- \imath \vec{k} \dot \vec{x}} T_{\mu \nu}
(\vec{x},t)~.
\end{eqnarray}
Thus, the gravitational power radiated per unit solid angle is
proportional to $T_{(p,q)}^2$, with $T_{(p,q)}$ given by
Eq.~(\ref{pq-gr-rad}). Hence, it is the same as that radiated by a
network of F-strings and a network of D-strings considered separately,
in analogy with the result found for cosmic strings in a
junction~\cite{Brandenberger:2008ni}.

\subsection{Axionic radiation}
Because $(p,q)$ strings are charged under $B_2^{\rm NS}$ and $C_2^{\rm
  RR}$, the NS-NS and Ramond-Ramond (RR) 2-forms, it should be
possible for them to lose energy via emission of massless RR or NS-NS
particles. The RR particles are often referred to as axions, because
the RR two-form $C_2$ is Hodge dual in four dimensions to a
pseudo-scalar known as the axion.

\subsubsection{RR particle emission}
In flat space, the power radiation via RR particle emission by
D-strings is comparable to the radiation by emission of gravitational
waves. However, in a warped background,
only the DBI
part of the action couples to the warp factor, so that it is possible
for energy loss via RR radiation to dominate over energy loss
by gravitational radiation \cite{Firouzjahi:2007dp}. We would like to
check this result for a general network of $(p,q)$ strings.

Considering axion emission from D-strings,
Ref.~\cite{Firouzjahi:2007dp} gives a comparison between the
gravitational and axionic radiation arising from the IIB D-string
action in four space-time dimensions
\begin{eqnarray}
S_{\rm D, 4dim} & = & \frac{M_{\rm P}^2}{2} \int d^4 x \sqrt{- g} \left ( R
- \frac{\beta g_s}{12} F_3^2\right ) - \mu_{eff} \int dt dx \sqrt{-
  \gamma} + \mu_1 \int dt dx C_2~,
\end{eqnarray}
where $g, \gamma, R, F_3^2, M_P$ are all four-dimensional quantities,
$\gamma_{ab}$ is the pull-back of the four-dimensional metric on the
string world-sheet (with the warp factor pulled out), and
\begin{eqnarray}
\mu_{\rm eff} & = & h^2 \mu_1 g_{\rm s}^{-1}~,\\
M_{\rm P}^2 & = & \frac{1}{\kappa_{10}^2} \int d^6 y \sqrt{g_6} h^2 (y)~,
\end{eqnarray}
with $\mu_1$ the string charge, and $\mu_1 = (2 \pi \alpha')^{-1}$.  Using
well-known results~\cite{Vachaspati:1984gt,Vilenkin:1986ku} for field
theory cosmic strings, one finds~\cite{Firouzjahi:2007dp}
\begin{eqnarray}
\label{power-RR-g}
P_{\rm RR} & = & \frac{\Gamma_{\rm RR} \mu_1^2}{ \pi^2 g_{\rm s} \beta
  M_{\rm P}^2}\\ P_{\rm g} & = &\Gamma_{\rm g} G \Big(\frac{h^2
  \mu_1}{g_{\rm s}}\Big)^2 ~,
\end{eqnarray}
leading to the ratio
\begin{eqnarray}
\label{RRratio}
\frac{P_{\rm RR}}{P_{\rm g}} & = & \Big(\frac{8\Gamma_{\rm RR}}{\pi
  \Gamma_{\rm g}} \Big ) \frac{g_{\rm s}}{\beta h^4}~,
\end{eqnarray}
where we have used the fact that $8 \pi G = M_P^{-2}$.  Note that $\Gamma_{\rm
  RR}$ and $\Gamma_{\rm g}$ are numerical factors of the same order
${\cal O}(50)$, and $\beta$ parametrizes the difference in
normalizations between the Chern-Simons and the Einstein-Hilbert term
in the presence of warping:
\begin{eqnarray}
\beta & = & \frac{\int d^6y\sqrt{g_6} h^{-2}(y)}{\int d^6 y \sqrt{g_6}
  h^2 (y)}~.
\end{eqnarray}
Thus, we see that in the limit where $g_{\rm s} \ll 1$ and warping is
negligible ({\sl i.e.}, $\beta \approx 1$), power loss by gravitational
radiation is dominant. However in a warped geometry, for instance a
Klebanov-Strassler background, $h = e^{2 \pi K/(3 g_{\rm s} M)}$ can
be much less than 1 and $P_{\rm RR}$ can be dominant; note that $K$
and $M$ are integers specifying the flux background, namely
\begin{equation}
(4 \pi^2 \alpha')^{-1} \int_B H_3 = - K \ \ \ \ \mbox {and}\ \ \ \ (4 \pi^2
\alpha')^{-1} \int_{S^3} F_3 = M~;
\end{equation}
the RR flux $F_3$ wraps the $S^3$ while the NS-NS flux $H_3$ wraps the
 Poincar\'{e} dual 3-cycle $B$ that generates the warped throat.

\subsubsection{NS particle emission}
In the case of F-strings, a kinetic term for $B_2$ must be
included. It appears in the ten-dimensional (Einstein-frame) action as
follows~\footnote{The IIB action can also be given in $SL(2,
  \mathbb{Z})$ invariant form in terms of $G_3 = F_3 - \tau H_3$
  where $\tau = C_0 + ie^{-\phi}$, but we consider the D- and F-string
  cases separately here.}:
\begin{eqnarray}
S_{\rm IIB} & = & \frac{1}{2 \kappa_{10}^2} \int d^{10}x \sqrt {-G} \left
[ R - \frac{1}{12 g_s} H_{(3)}^2 \right ] + S_{\rm local}~,
\end{eqnarray}
where $H = dB$ and  the $S_{\rm local}$ part of the action is given by
\begin{eqnarray}
S_{\rm local} & = & - \mu_1 \int d^2 \sigma\sqrt{- |\gamma_{ab}|} +
\mu_1 \int d^2 \sigma B_{2}^{\rm NS}~.
\end{eqnarray}
Following Ref.~\cite{Firouzjahi:2007dp} we compare the
four-dimensional action to that in Ref.~\cite{Vilenkin:1986ku},
finding (we have to take $B_2 \rightarrow \tilde B_2 = (M_P/2)
\sqrt{\beta/g_s} B_2$):
\begin{eqnarray}
\label{power-NS}
P_{\rm NS-NS} & = & \frac{\Gamma_{NS} \mu_1^2 g_{\rm s}}{ \pi^2 \beta
  M_{\rm P}^2}~.
\end{eqnarray}
Then from Eqs.~(\ref{power-RR-g}) and (\ref{power-NS}), the ratio $P_{\rm
  NS}/P_{\rm g}$ goes like $g_{\rm s}^3$:
  \begin{eqnarray}
  \label{ratioNS}
 \frac{P_{\rm NSNS}}{P_{\rm g}} & = & \Big(\frac{8\Gamma_{\rm NSNS}}{\pi
  \Gamma_{\rm g}} \Big ) \frac{g_{\rm s}^3}{\beta h^4}~, 
  \end{eqnarray}
  so this radiation is
suppressed compared to the RR particle radiation.\\

One might wonder what these ratios, Eqs.~(\ref{RRratio}) and
(\ref{ratioNS}), would be for $(p,q)$ strings, since if it is possible
for particle radiation to be dominant over graviton emission, this
would give an important observable difference between cosmic
superstrings and gauge strings, at least in the case where they are
produced in a warped throat. However, as we will show below, the
warped throat construction results in severe constraints on the
allowed radiation. In what follows, we will examine the validity of
the result claimed in Ref.~\cite{Firouzjahi:2007dp}, and its possible
extension in the case of $(p,q)$ strings, taking into account the
constraints from the orientifold projection imposed by a consistent
flux compactification.

\subsubsection{Constraints from the orientifold projection}
The enhancement of RR particle emission claimed for
D-strings~\cite{Firouzjahi:2007dp} is due to the effect of warping.
We reviewed above the construction and tension of $(p,q)$ strings in
a Klebanov-Strassler throat given in
Ref.~\cite{Firouzjahi:2006vp}. This is the relevant construction to
study if we want to answer the question of whether the enhancement of
RR radiation observed in Ref.~\cite{Firouzjahi:2007dp} for D-strings
carries over to the case of $(p,q)$ strings.  However, a careful
consideration of the throat geometry reveals some subtleties with
the construction. The only known consistent compactification of the
Klebanov-Strassler geometry is the flux compactification given by
Giddings, Kachru and Polchinski (GKP)~\cite{GKP} ({\sl see}, also
Ref.~\cite{Dasgupta:1999ss}), which involves an orientifold
projection~\footnote{Here we consider the orientifold projection
  corresponding to inclusion of ${\cal O}3$ and ${\cal O}7$-planes,
  which is the case for KS. Another possible orientifold,
  corresponding to the case of ${\cal O}5$ and ${\cal O}9$ planes,
  has orientifold action $\sigma^\star \Omega= +\Omega$. This would
  result in a different spectrum, since $C_2$ in the non-compact
  directions is not projected out in this case. However, at present
  no brane inflation model in a consistent compactification
  corresponding to this orientifold case is known to
  us. Furthermore, we remain at the orientifold limit, corresponding
  to the constant dilaton case. More complicated F-theory
  compactifications, in which the configuration of the ${\cal O}7$
  planes and D7-planes can vary, are possible for the non-constant
  $\tau$ case~\cite{Dasgupta:1999ss, GKP}, but we do not consider
  this case here.  We expect that any states which are able to
  survive the orientifold projection in this case will be
  massive~\cite{Dasgupta:2002ew, Sen:1998rg}, so they will not affect
  our argument.}
\begin{equation}
{\cal O} =  (-1)^{F_{\rm L}} \Omega_p \sigma^\star~,
\end{equation}
where $\sigma^\star$ is the pull-back of an isometric and holomorphic
involution $\sigma$ which leaves the K\"ahler form $J$ invariant
($\sigma^\star J=J$) but acts non-trivially on the holomorphic
three-form $\Omega$ (for ${\cal O}3/{\cal O}7$-orientifold planes
$\sigma^\star \Omega = - \Omega$), $\Omega_P$ is the world-sheet
parity and $F_L$ is the space-time fermion number in the left-moving
sector. [We refer the reader to, for example, Refs.~\cite{Grimm:2004uq,
    Cicoli:2009zh} for a detailed explanation of the orientifold
  action.] The action of ${\cal O}$ on the NS-NS and RR two-forms is
\begin{eqnarray}
{\cal O} B_2 \, \, =\, \, - \sigma^\star B_2
&&\ \ \mbox{and}\ \ \ \ {\cal O} C_2 \, \, = \, \, - \sigma^\star
C_2~,
\end{eqnarray}
respectively.  Since $\sigma$ is an internal symmetry which acts on
the internal manifold and leaves the four-dimensional non-compact
space invariant, the NS-NS and RR two-forms with legs in the
non-compact directions are projected out~\cite{Grana:2003ek}. This
means that there can be no massless RR or NS-NS axion, since the zero
modes of $B_{\mu \nu}$ and $C_{\mu \nu}$ do not appear in the
spectrum~\cite{CMP}. At this point it is worth noting that although
this implies that D, F or $(p,q)$ strings will not saturate the
Bogomol'nyi-Prasad-Sommerfeld (BPS) bound, because there is no gauge
group for them to be charged under, they will be stable against
annihilation with their image under the orientifold because this
annihilation is suppressed by the warping: the strings stretching
between them are massive~\cite{CMP}.  

One might thus wonder if axionic radiation is possible at all. To
check this, we decompose a general two-form:
\begin{eqnarray}
D_2 (x ) & = & d_2 (x^\mu) + d(x) \Omega_2 + V_1(x) \wedge \alpha_1~,
\end{eqnarray}
where $d(x)$ is a scalar, $d_2(x^ \mu)$ is a two-form in the
non-compact directions, $\Omega_2$ is a two-form in the internal
directions, and $\alpha_1$ a one-form in the internal
directions. While $d_2(x^\mu)$ is projected out by the orientifold
action, $\Omega_2$ is not as long as it is in the $-1$ eigenspace of
the involution $\sigma^\star$ in the orientifold projection. Because
${\cal O} D_2 = - \sigma^\star D_2$, this component of the two-form
will survive the orientifold projection, and indeed $d(x)$ is known
in the literature as a model-dependent axion \cite{Svrcek:2006yi}.
For a D-string charged under $C_{01}$ this makes no difference, as
$C_{01}$ is projected out. Since there is no massless RR mode, there
can be no axionic RR radiation from a D-string in a warped
background, so it does not make sense to compare this radiation to
the gravitational radiation from a D-string in a throat. As a
consequence, the result found in Ref.~\cite{Firouzjahi:2007dp} is
inapplicable.  Similarly, an F-string charged under $B_{01}$ will
have no axionic zero modes in a consistent warped geometry.  In
conclusion, neither D- nor F-strings can lead to significant axionic
emission, since by construction of the consistent brane inflationary
model that will lead to their formation, such objects will not have
any massless axionic radiation. 

However, for a $(p,q)$ string which is actually a wrapped D3-brane
with fluxes, as in Ref.~\cite{Firouzjahi:2006vp}, the situation may be
different.  The model-dependent axion $d(x)$ described above is
possible when $D_2$ has internal legs. Here $C_{\theta \phi}$ (or
$C_{23}$) is allowed, and couples to the string, as long as the
internal 2-cycle is odd under $\sigma^\star$. There is no such axion
for the NS-NS two-form, which must be along the $0,1$ directions, so
NS-NS particle radiation is entirely ruled out.  
  
For the $(p,q)$ string arising from a wrapped D3-brane, we should also  
consider an analogous decomposition for the RR four-form $C_4$:  
\begin{equation}  
C_4(x^M) = c_4 (x^\mu) + c_2 (x^\mu) \wedge \Omega_2 +  
  c_1(x^\mu)\wedge \Omega_3 + c(x^\mu) \wedge \Omega_4  
\ \ \ \ \mbox  
  {with}  \ \ \ \ {\cal O} C_4 = \sigma^\star C_4~.  
\end{equation}  
In our case, in which the D3-brane is wrapped on a 2-cycle, it is the  
term $c_2 (x^\mu) \wedge \Omega_2$ which is of interest, and which is  
allowed as long as $\Omega_2$ is odd under the involution.   
  
Harmonic $(p,q)$ forms are in one-to-one correspondence with the  
elements of the cohomology groups $H^{(p,q)}$ which split into two  
eigenstates under the action of $\sigma^\star$:  
\begin{eqnarray*}  
H^{(p,q)} & = & H_{+}^{(p,q)} \oplus H_{-}^{(p,q)}.  
\end{eqnarray*}  
The $\pm$ subscripts refer to even/odd behavior under  
$\sigma^\star$. Thus, in order for any RR mode to survive, it must be  
of the form $C_2 \wedge F_2$ on the D3-brane, where $F_2$ has legs in  
the $\theta$- and $\phi$-directions on a two cycle $\Omega_2 \in  
H_{-}^{(1,1)}$.       
  
Such a cycle is certainly allowed for a general $CY_3$. There is only  
one two-form on the $S^2$ wrapped by the D3-brane, the volume form  
$\omega$ which can be written as  
$$ \omega = x_1 d x_2 \wedge dx_3 + x_2 dx_3 \wedge d x_1 + x_3 dx_1  
\wedge dx_2$$ and is odd under, e.g., $x_1 \leftrightarrow x_2$. As  
long as such an involution is holomorphic and isometric on the full  
CY, these two modes are allowed.     
\subsubsection{The axionic wave-function}  
To determine allowed modes for radiation by cosmic superstrings in a  
warped geometry, it is necessary not only to take into account which  
modes survive the orientifold projection, but also to consider the  
correct dimensional reduction of these modes, which is a non-trivial  
task.  In Ref.~\cite{Frey:2008xw}, the equations of motion including  
warping (which includes compensating terms) were given for the  
universal K\"ahler (volume) modulus and its axionic partner $a$, the  
universal axion, which arises from the four-form as  
\begin{eqnarray}  
C_4 & = & a J \wedge J~,  
\end{eqnarray}  
where $J$ is the K\"ahler form. The action of the orientifold on  
$C_4$ is just ${\cal O} C_4 = \sigma^\star C_4$ and the involution  
$\sigma^\star$ leaves the K\"ahler form invariant, so this universal  
axion is allowed in a compactified throat geometry. However, the legs  
of the four-form are all in the internal manifold, so this axion  
cannot be sourced by the wrapped D3-brane. Note that the problem is  
not solved by wrapping a D5-brane (say) on a 4 cycle, because this 4  
cycle would have to be odd under the action of $\sigma^\star$ and  
cannot therefore be given by $J \wedge J$.    
  
Thus the only allowed axions are the non-universal axions. We have  
seen that only these could couple to the $(p,q)$ string constructed  
from a wrapped D3-brane.    
  
For the axion arising from $C_4 = c_2 (x) \wedge \Omega_2$, the  
analysis used in Ref.~\cite{Frey:2008xw} should be applicable, but it  
is not known how to solve the equations of motion for this case. An  
added complication is the fact that the compensators needed, mentioned  
in Section~\ref{warping}, result in mixing between $C_2$ and  
$C_4$. This is quite generic in dimensional reduction on warped  
geometries, because the backgrounds break diffeomorphism invariance,  
so that the correct gauge-invariant object to consider is a  
combination of different fields Ref.~\cite{Underwood:2010pm}. The  
resulting wave-function would affect the magnitude of any radiation in  
this mode, so without it it is not possible to quantify the amplitude  
of the radiation.
  
\subsection{Dilaton radiation} 
  
Let us briefly mention the possibility of dilatonic radiation
by cosmic superstrings. As noted in Ref.~\cite{Firouzjahi:2007dp}, the
dilaton is expected to be massive, making dilaton radiation by cosmic
superstrings negligible compared to the massless axionic radiation
modes. The dilaton mass is constrained by cosmological and
astrophysical observations ({\sl see e.g.},
Refs.~\cite{Tseytlin:1992jq, Babichev:2005qd}), and is
compatible with observations, only if it acquires a VEV early in
the history of the universe (before the end of
nucleosynthesis). Indeed, the dilaton gets a non-trivial potential in
the GKP compactification, because it enters the action in combination
with the RR and NS-NS fluxes as $G_3 = F_3 - \tau H_3$, where $\tau =
C_0 + \imath e^{- \phi}$, and $G_3$ must be imaginary selfdual.
However, as was shown in, {\sl e.g.}, Refs.~\cite{Burgess:2006mn,
  Frey:2006wv}, the mass of a dilaton in the throat will be suppressed
by the warp factor, namely
\begin{eqnarray*}  
m & \sim & \frac{e^{A}}{\sqrt{\alpha'}}~,
\end{eqnarray*}  
deep in the throat; with $A$ denoting the warp factor. It is therefore
conceivable that dilatons could be produced, particularly by high
energy processes like the decay of metastable strings via monopole
production \cite{Leblond:2009fq} .~\footnote{For it to make sense to
  consider only the lightest dilaton mode, one needs to check that the
  lowest mass is much smaller than the KK mass gap. Otherwise this
  lowest mass mode should be integrated out~\cite{Frey:2006wv}. The
  KKLT throat is a ``short throat" in the terminology of
  Ref.~\cite{Frey:2006wv}, which means that the mass gap allows one to
  keep the lowest mass dilaton mode, while the suppression of the mass
  compared to the unwarped case still holds.}
  
Constraints on cosmic superstring tension as a function of the dilaton
mass were obtained in Ref.~\cite{Babichev:2005qd}.  In
Ref.~\cite{Sabancilar} it was shown that in a warped geometry, these
constraints are weakened because, not only is the mass suppressed by
the warp factor, but the dilaton wavefunction will be localised in the
throat, with an exponential fall off in the bulk, which will increase
the strength of the coupling $\alpha$ of the dilaton to
matter~\cite{DeWolfe-Giddings, Giddings-Maharana, Burgess:2006mn, Frey:2006wv,
  Frey:2009qb}.~\footnote{There is a possible complication here:
  axion-dilaton fluctuations in a warped background will mix with
  metric fluctuations~\cite{Shiu:2008ry}. This will affect the
  wave-function. According to Ref.~\cite{Frey:2006wv} though, the
  contribution of compensators to the axion-dilaton mass will be
  negligible as long as the throat is long enough.} Dilaton radiation
from cosmic strings was studied in
Refs.~\cite{Damour:1996pv},~\cite{Peloso:2002rx},~\cite{Babichev:2005qd}
via the coupling of the dilaton field to matter fields forming the
strings. The interaction term in the Lagrangian taking the form
\begin{equation}  
\mathcal{L}_{int}\sim\frac{\alpha}{M_{Pl}}\phi T^\mu_\mu~,  
\end{equation}  
with $\alpha$ the coupling constant, the power of dilaton radiation  
was found to be proportional to the square of the coupling:  
\begin{equation}\label{dil-power}  
P_d\sim \Gamma\alpha^2 G\mu^2,  
\end{equation}  
where $\Gamma$ is a numerical factor of order 30. In the case of F-
and D-strings the constraints on the string tension from
dilaton emission can be further weakened depending on whether they
couple to matter or not~\cite{Sabancilar}.
\section{Conclusions}   
In this article we have set out to examine theoretical constraints on  
brane inflation models and the radiation channels of cosmic  
superstrings formed at the end of brane inflation. Cosmic superstrings  
can be produced at the end of brane inflation, whether it be D3/D7  
inflation which takes place in an unwarped geometry, or  
brane-antibrane inflation (for instance D3/$\overline{\rm D3}$)  
which takes place in a throat. Cosmic superstrings can arise in many  
different forms in string theory, as F-, D-, or FD-strings (or wrapped higher-dimensional branes) and  
correspondingly have more possible radiation modes than gauge theory  
cosmic strings. However, the constraints on both the brane  
inflationary models leading to cosmic superstring formation, as well as  
the radiation modes of cosmic superstrings coming from consistency of  
the string theory embedding, are quite dramatic.  
  
In the first part of this article we considered the implications of  
recent SUGRA constraints on models of brane inflation. It is necessary  
to check carefully that such models are consistent with these  
constraints, since in general in brane inflation models it is  
desirable to leave the brane positions unfixed while the volume of the  
compactification manifold is stabilized. In the case of D-term  
inflation models, a constant (field-independent) FI term is  
inconsistent with the SUGRA constraints, so these must be analyzed  
carefully.  
  
We found that D3/$\overline{\rm D3}$ inflation is consistent with the  
SUGRA constraints since the volume modulus can be fixed below the SUSY breaking scale. In D3/D7 inflation,  
which relies on the existence of an FI term, it is necessary for  
consistency of the model that the FI term depend on a modulus which is  
unfixed. We argue that this is the case -- that the FI term depends on  
the separation between the branes -- but point out that such a  
dependence may have implications for the inflationary dynamics in this  
model.  
In the second part of the article we focussed on radiation channels of  
cosmic superstrings in a warped background, which are also subject to  
strong constraints. It is in warped throats that cosmic superstring  
tension can be lowered to within observationally acceptable bounds  
(without the need for the strings to be semilocal, for  
instance). Furthermore, it was argued in Ref.~\cite{Firouzjahi:2007dp}  
that warping could result in enhancement of axionic radiation by  
D-strings as compared to gravitational radiation.  We have seen that  
the results of Ref.~\cite{Firouzjahi:2007dp} do not translate easily  
to the case of $(p,q)$ strings in a throat. To begin with, axionic  
radiation from either F- or D-strings is ruled out in a consistent  
warped compactification because the modes $B_2$ and $C_2$ are  
projected out by the orientifold action. Thus, it is not possible for  
axionic radiation from D-strings in a throat to be enhanced relative  
to gravitational radiation. For $(p,q)$ strings which are actually  
branes wrapped on suitable cycles ({\sl e.g.}, a D3-brane wrapped on  
$\Omega_2$ where $\Omega_2$ is odd under the involution  
$\sigma^\star$) axionic radiation is still possible in  
principle. However, the wave-function of the axionic zero mode is  
non-trivially modified by the warping, and only limited progress has  
been made in taking these modifications into account. The equations of  
motion for the universal axion can be written down, but the same is  
not yet true for the non-universal axion.  We do not see any way for a  
$(p,q)$ string in a throat to couple to the universal axion, which  
means that it is not currently possible to calculate correctly the  
power of the Ramond-Ramond radiation by these strings. 

\begin{acknowledgments}

It is a pleasure to thank Keshav Dasgupta, Andrew Frey, Zohar Komargodski,  Liam McAllister, Sakura
Schafer-Nameki and Bret Underwood for correspondence and
discussions.  This work is partially supported by the Sciences \&
Technology Facilities Council (STFC--UK), Particle Physics Division,
under the grant ST/G000476/1 ``Branes, Strings and Defects in
Cosmology''. The work of R.~G. is also supported by an NSERC
Postdoctoral Fellowship.
\end{acknowledgments}

\end{document}